\documentclass{article}
\pdfoutput=1
\usepackage{graphicx,xspace,url}
\usepackage{booktabs,color}
\usepackage{listings} 
\usepackage{trcover}

\trnum{TR-10-20}
\date{November 30, 2010}
\usepackage{makeidx,graphicx,booktabs,xspace,url,listings,color,multicol,pifont,multirow,textcomp,subfig,setspace,paralist}

\newtheorem{theorem}{Theorem}[section]

\lstnewenvironment{Bench}[2]
{\lstset{ %
showlines=true,
xleftmargin=5ex,
firstnumber=1,
language=C++,                % choose the language of the code
basicstyle=\scriptsize,       % the size of the fonts that are used for the code
numbers=left,                   % where to put the line-numbers
numberstyle=\tiny,      % the size of the fonts that are used for the line-numbers
stepnumber=1,                   % the step between two line-numbers.
numbersep=3pt,                  % how far the line-numbers are from the code
showspaces=false,               % show spaces adding particular underscores
showstringspaces=false,         % underline spaces within strings
showtabs=false,                 % show tabs within strings adding particular underscores
tabsize=2,                      % sets default tabsize to 2 spaces
captionpos=b,                   % sets the caption-position to bottom
breaklines=true,                % sets automatic line breaking
breakatwhitespace=false,        % sets if automatic breaks should only happen at whitespace
escapeinside={\%*}{*)},          % if you want to add a comment within your code
mathescape=true,
columns=flexible,
morekeywords={main,add_workers,add_emitter,wrap_around,
run_and_wait_end,ff_send_out},
#1%, label={code:#2}
}}{}

\lstnewenvironment{Bench2}[2]
{\lstset{ %
xleftmargin=5ex,
linewidth=1.1\linewidth,
firstnumber=1,
language=C++,                % choose the language of the code
multicols=2,
basicstyle=\scriptsize,       % the size of the fonts that are used for the code
numbers=left,                   % where to put the line-numbers
numberstyle=\tiny,      % the size of the fonts that are used for the line-numbers
stepnumber=1,                   % the step between two line-numbers.
numbersep=3pt,                  % how far the line-numbers are from the code
showspaces=false,               % show spaces adding particular underscores
showstringspaces=false,         % underline spaces within strings
showtabs=false,                 % show tabs within strings adding particular underscores
tabsize=2,                      % sets default tabsize to 2 spaces
captionpos=b,                   % sets the caption-position to bottom
breaklines=true,                % sets automatic line breaking
breakatwhitespace=false,        % sets if automatic breaks should only happen at whitespace
escapeinside={\%*}{*)},          % if you want to add a comment within your code
mathescape=true, columns=flexible, morekeywords={main,add_workers,add_emitter,wrap_around,run_and_wait_end,ff_send_out},
#1%, label={code:#2}
}}{}

\newcommand {\lines}[2]{\textsc{\small \S #1.\oldstylenums{#2}}}

\title{Single-Producer/\\Single-Consumer Queues\\on Shared Cache Multi-Core Systems}

\author{Massimo Torquati\\
Computer Science Department \\
University of Pisa, Italy. \\
Email: torquati@di.unipi.it}

\begin{document}
\maketitle
\begin{abstract}
Using efficient point-to-point communication channels is critical for 
implementing fine grained parallel program on modern shared cache multi-core
architectures.

This report discusses in detail several implementations of wait-free
Single-Producer/Single-Consumer queue (SPSC), and presents a novel 
and efficient algorithm for the implementation of an unbounded 
wait-free SPSC queue (uSPSC). The correctness proof of the new algorithm, and 
several performance measurements based on simple synthetic 
benchmark and microbenchmark, are also discussed.
 
\end{abstract}

\section{Introduction}

This report focuses on Producer-Consumer coordination, and in
particular on  Single-Producer/Single-Consumer (SPSC) coordination.
The producer and the consumer are concurrent entities, i.e. processes or threads.
The first one produces items placing them in a shared structure, whereas the 
the second one consumes these items by removing them from the shared structure.
Different kinds of shared data structures provide different fairness guarantees. 
Here, we consider a queue data structure that provides First-In-First-Out 
fairness (FIFO queue), and we assume that Producer and Consumer share a common
address space, that is, we assume threads as concurrent entities.

In the end of '70s, Leslie Lamport proved that, under Sequential Consistency 
memory model \cite{LamportSC}, a Single-Producer/Single-Consumer circular buffer \footnote{A circular buffer can be used to implement a FIFO queue} can be 
implemented without using explicit synchronization mechanisms between the 
producer and the consumer \cite{Lamport77}. Lamport's circular buffer is a 
wait-free algorithm. A wait-free algorithm is guaranteed to complete after a
finite number of steps, regardless of the timing behavior of other operations. 
Differently, a lock-free algorithm guarantees only that after a finite 
number of steps, \texttt{some} operation completes. Wait-freedom is a stronger 
condition than lock-freedom and both conditions are strong enough to preclude 
the use of blocking constructs such as locks.

With minimal modification to Lamport's wait-free
%%\footnote{A wait-free algorithm \cite{Herlihy91} is always lock-free, but not vice-versa} 
SPSC algorithm, it results correct also under 
Total-Store-Order and others weaker consistency models, but it fails under weakly 
ordered memory model such as those used in IBM's Power and Intel's Itanium
architectures. On such systems, expensive memory barrier (also known as memory fence)
instructions are needed in order to ensure correct load/store instructions ordering.

Maurice Herlihy in his seminal paper \cite{Herlihy91} formally proves that few 
simple HW atomic instructions are enough for building any wait-free
data structure for any number of concurrent entities. The simplest and widely 
used primitive is the compare-and-swap (CAS). 
Over the years, many works have been proposed with focus on lock-free/wait-free 
Multiple-Producer/Multiple-Consumer (MPMC) queue \cite{MNSS,mpmc1,ABA:98}.
They use CAS-like primitives in order to guarantee correct implementation.

Unfortunately, the CAS-like hardware primitives used in the implementations, 
introduce non-negligible overhead on modern shared-cache architectures, 
so even the best MPMC queue implementation, is not able to obtain better 
performance than Lamport's circular buffer in cases with just 1 producer and 
1 consumer.

FIFO queues are typically used to implement streaming networks \cite{fastflow:web, streamIt}. Streams are directional 
channels of communication that behave as a FIFO queue. In many cases streams are 
implemented using circular buffer instead of a pointer-based dynamic queue in order to avoid excessive memory usage. Hoverer, when 
complex streaming networks have to be implemented, which have multiple nested cycles, 
the use of bounded-size queues as basic data structure requires more complex and 
costly communication prtocols in order to avoid deadlock situations.

Unbounded size queue are particularly interesting in these complex cases, 
and in all the cases where it is extremely difficult to choose a suitable 
queue size. As we shall see, it is possible to implement a wait-free unbounded 
SPSC queue by using Lamport's algorithm and dynamic memory allocation. 
Unfortunately, dynamic memory allocation/deallocation is costly because they use locks 
to protect internal data structures, hence introduces costly memory barriers.

In this report it is presented an efficient implementation of an unbounded wait-free 
SPSC FIFO queue which makes use only of a modified version of the Lamport's circular
buffer without requiring any additional memory barrier, and, at the same time,
minimizes the use of dynamic memory allocation. 
The novel unbounded queue implementation presented here, is able to speed up producer-consumer 
coordination, and, in turn, provides the basic mechanisms for implementing complex streaming 
networks of cooperating entities. 
 
The remainder of this paper is organized as follows. Section \ref{sec:SWSR}
reminds Lamport's algorithm and also shows the necessary modifications to
make it work efficiently on modern shared-cache multiprocessors. 
Section \ref{sec:basicUnbounded} discuss the extension of the Lamport's algorithm 
to the unbounded case. Section \ref{sec:uSWSR} presents the new implementations
with a proof of correctness. Section \ref{sec:exp} presents some
performance results, and Sec. \ref{sec:conclusions} concludes.

\section{Lamport's circular buffer}
\label{sec:SWSR}

\begin{figure}
\begin{Bench2}{}{}
bool push(data) {
  if ((tail+1 mod N)==head ) 
    return false; // buffer full
  buffer[tail]=data;
  tail= tail+1 mod N;
  return true;
}
bool pop(data) {
  if (head==tail) 
    return false; // buffer empty
  data = buffer[head];
  head = head+1 mod N;
  return true;
}
\end{Bench2}
\caption{Lamport's circular buffer \texttt{push} and \texttt{pop} methods pseudo-code.
At the beginning \texttt{head}=\texttt{tail}=0. \label{fig:Lamport}}
\hspace{0.5ex}
\end{figure}

\begin{figure}
\begin{Bench2}{}{}
bool push(data) {
  if (buffer[tail]==BOTTOM) { 
    buffer[tail]=data;
    tail = tail+1 mod N;
    return true;
  }
  return false; // buffer full
}

bool pop(data) {
  if (buffer[head]!=BOTTOM) {
    data = buffer[head];
    buffer[head] = BOTTOM;
    head = head+1 mod N;
    return true;
  }
  return false; // buffer empty
}
\end{Bench2}
\caption{$P_1C_1$-buffer buffer pseudocode. Modified version of the code presented in \cite{HK97}. The buffer is initialized to BOTTOM and \texttt{head}=\texttt{tail}=0 at the beginning.\label{fig:P1C1buffer}}
%%\hspace{0.5ex}
\end{figure}

In Fig. \ref{fig:Lamport} the pseudocode of the
\texttt{push} and \texttt{pop} methods 
of the Lamport's circular buffer algorithm, is sketched. The \texttt{buffer} is implemented
as an array of \texttt{N} entries.

Lamport proved that, under Sequential Consistency \cite{LamportSC}, no locks are
needed around \texttt{pop} and \texttt{push} methods, thus resulting
in a concurrent wait-free queue implementation.
If Sequential Consistency requirement is released, it is easy to see that Lamport's 
algorithm fails. This happens for example with the  
PowerPC architecture where write to write relaxation is allowed ($W \rightarrow W$ using the 
same notation used in \cite{Adve95sharedmemory}),  
i.e. 2 distinct writes at different memory locations may be executed
not in program order. In fact, the consumer may pop out of the  
buffer a value before the data is effectively written in it, this is because 
the update of the \texttt{tail} pointer (modified only by the producer) can be seen 
by the consumer before the producer writes in the \texttt{tail} position of the buffer.
In this case, the test at line \lines{\ref{fig:Lamport}}{9} would be passed even though 
\texttt{buffer[head]} contains stale data.

Few simple modifications to the basic Lamport's algorithm, allow the correct 
execution even under weakly ordered memory consistency model. 
To the best of our knowledge such modifications have been presented and 
formally proved correct for the first time by Higham and Kavalsh in \cite{HK97}.
The idea mainly consists in tightly coupling control and data information into a 
single buffer operation by using a know value (called BOTTOM), which cannot be used by
the application. The BOTTOM value is used to indicate whether there is an empty buffer slot, 
which in turn indicates an available room in the buffer to the producer and the empty 
buffer condition to the consumer. 

With the circular buffer implementation sketched in Fig. \ref{fig:P1C1buffer}, the consistency 
problem described for the Lamport's algorithm cannot occur provided that the generic store 
\texttt{buffer[i]=data} is seen in its entirety by a processor, or not at all, i.e. 
a single memory store operation is executed atomically. To the best of our knowledge, 
this condition is satisfied in any modern general-purpose processor for aligned memory word stores.

As shown by Giacomoni et all. in \cite{fastforward:ppopp:08}, Lamport's circular buffer algorithm
results in cache line thrashing on shared-cache multiprocessors, as the \texttt{head} and 
\texttt{tail} buffer pointers are shared between consumer and producer. 
Modifications of pointers, at lines \lines{\ref{fig:Lamport}}{5} and \lines{\ref{fig:Lamport}}{12}, turn out 
in cache-line invalidation (or update) traffic among processors, thus introducing unexpected overhead.
With the implementation in Fig. \ref{fig:P1C1buffer}, the head and the \texttt{tail} buffer pointers are
always in the local cache of the consumer and the producer respectively, without incurring in 
cache-coherence overhead since they are not shared. 

When transferring references through the buffer rather than plain data values, a memory fence is required
on processors with weakly memory consistency model, in which stores can be executed out of 
program order. In fact, without a memory fence, the write of the reference in the buffer 
could be visible to the consumer before the referenced data has been committed in memory. 
In the code in Fig. \ref{fig:P1C1buffer}, a write-memory-barrier (WMB) must be inserted between
line \lines{\ref{fig:Lamport}}{2} and line \lines{\ref{fig:Lamport}}{3}.

\begin{figure}
\begin{Bench}{}{}
class SPSC_buffer {
private:
  volatile unsigned long    pread;
  long padding1[longxCacheLine-1];
  volatile unsigned long    pwrite;
  long padding2[longxCacheLine-1];
  const    size_t           size;
  void                   ** buf;
public:
  SWSR_Ptr_Buffer(size_t n, const bool=true):
     pread(0),pwrite(0),size(n),buf(0) {
  }    
  ~SWSR_Ptr_Buffer() { if (buf)::free(buf); }

  bool init() {
     if (buf) return false;
       buf = (void **)::malloc(size*sizeof(void*));
       if (!buf) return false;
       bzero(buf,size*sizeof(void*));
       return true;
  }

  bool empty()     { return (buf[pread]==NULL);}
  bool available() { return (buf[pwrite]==NULL);}

  bool push(void * const data) {
     if (available()) {
       WMB(); 
       buf[pwrite] = data;
       pwrite += (pwrite+1 >= size) ? (1-size): 1;
       return true;
     }
     return false;
  }
  bool  pop(void ** data) {
     if (empty()) return false;        
     *data = buf[pread];
     buf[pread]=NULL;
     pread += (pread+1 >= size) ? (1-size): 1;        
     return true;
  }    
};
\end{Bench}
\caption{SPSC circular buffer implementation.\label{fig:SPSC}}
%%\hspace{0.5ex}
\end{figure}

The complete code of the SPSC circular buffer is shown in Fig. \ref{fig:SPSC}.

\subsection{Cache optimizations}
\label{sec:cacheopt}

Avoiding cache-line thrashing due to false-sharing is a critical aspect in shared-cache multiprocessors.
Consider the case where two threads sharing a SPSC buffer are working in lock step.
The producer produces one task at a time while the consumer immediately consumes 
the task in the buffer.
When a buffer entry is accessed, the system reads a portion of memory containing
the data being accessed placing it in a cache line. 
The cache line containing the buffer entry is read by the consumer thread
which only consumes one single task. The producer than produces the next task pushing
the task into a subsequent entry into the buffer. Since, in general, a single cache 
line contains several buffer entries (a typical cache line is 64bytes, whereas a
memory pointer on a 64bit architecture is 8 bytes) the producer's 
write operation changes the cache line status invalidating the whole
contents in the line. 
When the consumer tries to consume the next task the entire cache line is reloaded 
again even if the consumer tries to access a different buffer location.
This way, during the entire computation the cache lines containing the buffer entries 
bounce between the producer and the consumer private caches incurring in extra overhead 
due to cache coherence traffic. 
The problem arises because the cache coherence protocol works at cache
line granularity and because the ``distance'' between the producer and the consumer (i.e. $|pwrite-pread|$) 
is less than or equal to the number of tasks which fill a cache line (on a 64bit machine with 64bytes
of cache line size the critical distance is 8).
In order to avoid false sharing between the head and tail pointers in the SPSC queue, a proper amount of padding in required to force the two pointers to reside in different cache lines (see for example Fig. \ref{fig:SPSC}).
In general, the thrashing behavior can be alleviated if the producer and the consumer are 
forced to work on different cache lines, that is, augmenting the ``distance''.

The FastForward SPSC queue implementation presented in
\cite{fastforward:ppopp:08}  improves 
Lamport's circular buffer implementation by optimizing cache behavior and preventing 
cache line thrashing.
FastForward temporally slips the producer and the consumer in such a way that 
push and pop methods operate on different cache lines. The consumer,
upon receiving its first task, spins until an appropriate amount of slip 
(that is the number of tasks in the queue reach a fixed value) is established. 
During the computation, if necessary, the temporal slipping is maintained by the 
consumer through local spinning.
FastForward obtains a performance improvement of 3.7 over Lamport's circular buffer
when temporal slipping optimization is used.

A different approach named cache line protection has been used in MCRingBuffer 
\cite{MCRINGBUFFER:ipdps:10}.
The producer and consumer thread update private copies of the head and tail buffer
pointer for several iterations before updating a shared copy. Furthermore, MCRingBuffer
performs batch update of control variables thus reducing the frequency of writing the shared
control variables to main memory.

A variation of the MCRingBuffer approach is used in Liberty Queue \cite{LQ:10}. Liberty
Queue shifts most of the overhead to the consumer end of the queue. Such customization 
is useful in situations where the producer is expected to be slower
than the consumer.

\begin{figure}
\begin{Bench}{}{}
bool multipush(void * const data[], int len) {        
  unsigned long last = pwrite + ((pwrite+ --len >= size) ? (len-size): len);
  unsigned long r    = len-(last+1), l=last, i;
  if (buf[last]==NULL) {          
    if (last < pwrite) {
      for(i=len;i>r;--i,--l) 
        buf[l] = data[i];
      for(i=(size-1);i>=pwrite;--i,--r)
        buf[i] = data[r];
    } else 
      for(register int i=len;i>=0;--i) 
        buf[pwrite+i] = data[i];
          
    WMB();
    pwrite = (last+1 >= size) ? 0 : (last+1);
    mcnt = 0; // reset mpush counter
    return true;
  }
  return false;
}

bool flush() {
  return (mcnt ? multipush(multipush_buf,mcnt) : true);
}

bool mpush(void * const data) {
  if (mcnt==MULTIPUSH_BUFFER_SIZE)
    return multipush(multipush_buf,MULTIPUSH_BUFFER_SIZE);

  multipush_buf[mcnt++]=data;

  if (mcnt==MULTIPUSH_BUFFER_SIZE)
    return multipush(multipush_buf,MULTIPUSH_BUFFER_SIZE);

  return true;
}

\end{Bench}
\caption{Methods added to the SPSC buffer to reduce cache trashing.\label{fig:SPSC:multipush}}
\end{figure}

\textbf{Multipush method.} Here we present a sligtly different approach 
for reducing cache-line trashing which is very simple and effective, 
and does not introduce any significant modification to the basic SPSC queue 
implementation.
The basic idea is the following: instead of enqueuing just one item at a time 
directly into the SPSC buffer, we can enqueue the items in a temporary array 
and then submit the entire array of tasks in the buffer
using a proper insertion order.
We added a new method called \texttt{mpush} to the SPSC buffer implementation
(see Fig.\ref{fig:SPSC:multipush}),
which has the same interface of the push method but inserts the 
data items in a temporary buffer of fixed size. 
The elements in the buffer are written in the SPSC buffer 
only if the local buffer is full or if the \texttt{flush} method is called.
The \texttt{multipush} method gets in input an array of items, and 
writes the items into the SPSC buffer in backward order.
The backward order insertions, is particularly important to reduce 
cache trashing, in fact, in this way, we enforce a distance between 
the \texttt{pread} and the \texttt{pwrite} pointers thus reducing the cache invalidation ping-pong.
Furthermore, writing in backward order does not require any other 
control variables or synchronisation.

This simple approach increase cache locality by reducing the 
cache trashing. However, there may be two drawbacks:
\begin{enumerate}
\item we pay an extra copy for each element to push into the SPSC buffer
\item we could increase the latency of the computation if the consumer
is much faster than the producer.
\end {enumerate}

The first point, in reality, is not an issue because the cost of extra
copies are typically amortized by the better cache utilization. The
second point might represent an issue for applications exhibiting very
strict latency requirements that cannot be balanced by an increased
throughput (note however that this is a rare requirement in a streaming application).  
In section \ref{sec:exp}, we try to evaluate experimentally the benefits
of the proposed approach.

\section{Unbounded List-Based Wait-Free SPSC Queue}
\label{sec:basicUnbounded}

Using the same idea of the Lamport's circular buffer algorithm,
it is possible to implement an unbounded wait-free SPSC queue 
using a list-based algorithm and dynamic memory allocation/deallocation.
The implementation presented here has been inspired by the work
of Hendler and Shavit in \cite{Hendler02b}, although it is different in
several aspects.
The pseudocode is sketched in Fig. \ref{fig:listSPSC}.

\begin{figure}
\begin{Bench2}{}{}

bool push(data) {
  Node * n = allocnode(data);
  WMB();
  tail->next = n; 
  tail       = n;
  return true;
}

bool  pop(data) {        
  if (head->next != NULL) {
    Node * n = (Node *)head;
    data     = (head->next)->data;
    head     = head->next;
    deallocnode(n);
    return true;
  }
  return false; // queue empty
}    
\end{Bench2}
\caption{Unbounded list-based SPSC queue implementation.\label{fig:listSPSC}}
\hspace{0.5ex}
\end{figure}

The algorithm is very simple: the \texttt{push} method 
allocates a new \texttt{Node} data structure containing the real value
to push into the queue and a pointer to the next \texttt{Node}
structure. The tail pointer is adjusted to point to the current Node.
The \texttt{pop} method gets the current head Node, sets the data value, 
adjusts the head pointer and, before exiting, deallocates the head 
\texttt{Node}.

In the general case, the main problem with the list-based
implementation of queues is the dynamic
memory allocation/deallocation of the \texttt{Node} structure. In fact,
dynamic memory management  
operations, typically, use lock to enforce mutual exclusion to protect
internal shared data structures,  
so, much of the benefits gained using lock-free implementation of the
queue are eventually  
lost. %%when memory is dynamically allocated or released. 
To mitigate such overhead, it is possible to use caching of list's internal 
structure (e.g. \texttt{Node}) \cite{Hendler02b}. The cache is by definition
bounded in the number of  
elements and so it can be efficiently implemented using a wait-free
SPSC circular buffer 
presented in the previous sections. 
Figure \ref{fig:SPSC_dynBuffer} shows the complete implementation of the list-based SPSC queue when 
Node caching is used. In the following we will refer to this
implementation with the name dSPSC. 

\begin{figure}
\begin{Bench2}{}{}
class SPSC_dynBuffer {
  struct Node {
     void        * data;
     struct Node * next;
  };
  volatile Node *  head;
  long pad1[longxCacheLine-sizeof(Node *)];
  volatile Node *  tail;
  long pad2[longxCacheLine-sizeof(Node*)];
  SPSC_Buffer      cache;
private:
  Node * allocnode() {
    Node * n = NULL;
    if (cache.pop((void **)&n)) return n;
      n = (Node *)malloc(sizeof(Node));
      return n;
  }
public:
  SPSC_dynBuffer(int size):cache(size) {
    Node * n=(Node *)::malloc(sizeof(Node));
    n->data = NULL; n->next = NULL;
    head=tail=n;  cache.init();
  }

  ~SPSC_dynBuffer() { ... }

  bool push(void * const data) {
     Node * n = allocnode();
     n->data = data; n->next = NULL;
     WMB();
     tail->next = n;
     tail       = n;
     return true;
  }
  bool  pop(void ** data) {        
     if (head->next) {
       Node * n = (Node *)head;
       *data    = (head->next)->data;
       head     = head->next;
       if (!cache.push(n)) free(n);
       return true;
     }
     return false;
  }    
};
\end{Bench2}
\caption{Unbounded list-based SPSC queue implementation with Node(s) caching (dSPSC).\label{fig:SPSC_dynBuffer}}
%%\hspace{0.5ex}
\end{figure}

As we shall see in Sec. \ref{sec:exp}, caching strategies help in
improving the  performance but are not sufficient  
to obtain optimal figures. This is mainly due to the poor cache
locality caused by lots of memory 
indirections. Note that the number of elementary
instruction per push/pop operation 
is greater than the ones needed in the SPSC implementation.

%%Dire che anche in questo caso ci vuole una WMB \mt{}.
%%Poor cache locality due to indirection.
%Inoltre il numero di operazioni elementari per ogni push e pop sono piu' alte del caso bounded.

\section{Unbounded Wait-Free SPSC Queue}
\label{sec:uSWSR}

We now describe an implementation of the unbounded wait-free SPSC queue
combining the ideas described in the previous sections. 
We refer to the implementation with the name uSPSC.

The key idea is quite simple: the unbounded queue is based on a pool 
of wait-free SPSC circular buffers (see Sec. \ref{sec:SWSR}). The pool of buffers 
automatically grows and shrinks on demand. The implementation of the pool of
buffers carefully try to minimize the impact of dynamic memory allocation/deallocation 
by using caching techniques like in the list-based SPSC queue. 
Furthermore, the use of SPSC circular buffers as basic uSPSC data structure, 
enforce cache locality hence provides better performance.

The unbounded queue uses two pointers: buf\_w that points to writer's buffer 
(the same of the tail pointer in the circular buffer), and a buf\_r that points 
to reader's buffer (the same of the head pointer). 
Initially both buf\_w and buf\_r point to the same SPSC circular buffer.
The push method works as follow. The producer first checks whether there is an 
available room in the current buffer (line \lines{\ref{fig:uSPSC}}{52}) and then push the data.
If the current buffer is full, asks the pool for a new buffer (line \lines{\ref{fig:uSPSC}}{53}),
set the buf\_w pointer and push the data into the new buffer.

\begin{figure}
\begin{Bench2}{}{}
class BufferPool {
  SPSC_dynBuffer inuse;
  SPSC_Buffer    bufcache;
public:
  BufferPool(int cachesize)
   :inuse(cachesize),bufcache(cachesize) {
     bufcache.init();
  }    
  ~BufferPool() {...}
   
  SPSC_Buffer * const next_w(size_t size)  { 
    SPSC_Buffer * buf;
    if (!bufcache.pop(&buf)) {
      buf = new SPSC_Buffer(size);
      if (buf->init()<0) return NULL;
    }        
    inuse.push(buf);
    return buf;
  }
  SPSC_Buffer * const next_r()  { 
    SPSC_Buffer * buf; 
    return (inuse.pop(&buf)? buf : NULL);
  }
  void release(SPSC_Buffer * const buf) {
    buf->reset();
    if (!bufcache.push(buf)) delete buf;
  }    
};
class uSPSC_Buffer {
  SPSC_Buffer * buf_r;
  long padding1[longxCacheLine-1];
  SPSC_Buffer * buf_w;
  long padding2[longxCacheLine-1];
  size_t        size;
  BufferPool    pool;
public:
  uSPSC_Buffer(size_t n)
    :buf_r(0),buf_w(0),size(size),
     pool(CACHE_SIZE) {}    
  ~uSPSC_Buffer() { ... }
    
  bool init() {
    buf_r = new SPSC_Buffer(size);
    if (buf_r->init()<0) return false;
    buf_w = buf_r;
    return true;
  }
  bool empty() {return buf_r->empty();}
  bool available(){return buf_w->available();}
 
  bool push(void * const data) {
    if (!available()) {
      SPSC_Buffer * t = pool.next_w(size);
      if (!t) return false; 
      buf_w = t;
    }
    buf_w->push(data);
    return true;
  }    
  bool  pop(void ** data) {
    if (buf_r->empty()) {
      if (buf_r == buf_w) return false;
      if (buf_r->empty()) {
        SPSC_Buffer * tmp = pool.next_r();
        if (tmp) {
          pool.release(buf_r); 
          buf_r = tmp;                    
        }
      }
    }
    return buf_r->pop(data);
  }    
};
\end{Bench2}
\caption{Unbounded wait-free SPSC queue implementation.\label{fig:uSPSC}}
%%\hspace{0.5ex}
\end{figure}

The consumer first checks whether the current buffer is not empty and in case  
pops the data. If the current buffer is empty, there are 2 possibilities:
\begin{enumerate}
\item there are no items to consume, i.e. the unbounded queue is really empty;
\item the current buffer is empty (i.e. the one pointed by buf\_r), but there may 
be some items in the next buffer.
\end{enumerate}

For the consumer point of view, the queue is really empty when the current
buffer is empty and both the read and write pointers point to the same buffer.
If the read and writer queue pointers differ, the consumer have to re-check the 
current queue emptiness because in the meantime (i.e. between the execution of
the instruction \lines{\ref{fig:uSPSC}}{61} and \lines{\ref{fig:uSPSC}}{62}) 
the producer could have written some new elements into the current buffer before 
switching to a new one. This is the most subtle condition, if we switch to the
next buffer but the current one is not really empty, we definitely loose data.
If the queue is really empty for the consumer, the consumer switch to a new buffer  
releasing the current one in order to be recycled by the buffer pool
(lines \lines{\ref{fig:uSPSC}}{64}--\lines{\ref{fig:uSPSC}}{67}).

%% Now we have to see the implementation of the pool class and in particular
%% of the methods next_r, next_w and release.

%% pool
%% First we check whether there is one available buffer in the cache. 
%% The cache is just a SWSR queue, so no lock around push and pop operation
%% is required. If no queue is available in the cache, it is allocated 
%% a new buffer and then the poiter pushed into the pool of the inuse queues. 
%% The inuse pool is protected by spin\_lock calls, which are implemented using 
%% CAS operation (CompareAndExchange).
%% In this case it is necessary to use mutual exclusion because the inuse pool
%% data structure have to be unbounded and so cannot be implemented by using a 
%% SWSR lock-free queue. 

%% The next_r method just try to pop out of the inuse pool a new queue pointer.

%% The release method just try to push the queue pointer
%% into the cache, if the push method fails, it means that maximum cache size
%% has been reached and so we have to discard the pointer releasing the allocated memory.

\subsection{Correctness proof}

Here we provide a proof of correctness for the uSPSC queue implementation
described in the previous section. 
By correctness, we mean that the consumer extracts elements from 
the queue in the same order in which they were inserted by the
producer.
%%\ma{Questo e' un concetto sufficiente di correttezza? }
The proof is based on the only assumption that the SPSC circular buffer 
algorithm is correct. A formal proof of the correctness of the SPSC buffer 
can be found in \cite{HK97} and \cite{fastforward:ppopp:08}.
Furthermore, the previous assumption, implies that
memory word read and write are executed atomically. This is one 
of the main assumption for the proof of correctness for the 
SPSC wait-free circular buffer \cite{fastforward:ppopp:08}. 
To the best of our knowledge, this condition is satisfied in any modern CPUs.

The proof is straightforward. 
If buf\_r differs from buf\_w, the execution is correct because there is no data sharing
between producer and consumer (the push method uses only the
buf\_w pointer, whereas the pop method uses only the buf\_r pointer),
since the producer and the consumer use different SPSC buffer. 
If buf\_r is equal to  buf\_w (both the producer and the consumer use
the same SPSC buffer) and the buffer is neither seen full 
nor empty by the producer and the consumer, the execution is correct because
of the correctness of the SPSC circular buffer.
So, we have to prove that if buf\_r is equal to buf\_w and the buffer is 
seen full or empty by the producer and/or by the consumer respectively, 
the execution of the \texttt{push} and \texttt{pop} methods are always correct.

The previous sentence has only one subtle condition worth proving:
buf\_r is equal to buf\_w and the producer sees the buffer full whereas 
the consumer sees the buffer empty. This sound strange but it is not. 

Suppose that the internal SPSC buffers used in the implementation of the uSPSC queue
has only a single slot (size=1). Suppose also that the consumer try to pop one 
element out of the queue, and the queue is empty. 
Before checking the condition at line \lines{\ref{fig:uSPSC}}{62},
the producer inserts an item in the queue and try to insert a second one.
At the second insert operation, the producer gets a new buffer because
the current buffer is full (line \lines{\ref{fig:uSPSC}}{53}), so, the buf\_w 
pointer changes pointing to the new buffer (line \lines{\ref{fig:uSPSC}}{55}).
Since we have not assumed anything about read after write memory ordering 
($R \rightarrow W$ using the same notation as in \cite{Adve95sharedmemory}),  
we might suppose that the write of the buf\_w pointer is immediately visible to the consumer 
end such that for the consumer results buf\_r different from buf\_w at line \lines{\ref{fig:uSPSC}}{62}.
In this case, if the consumer sees the buffer empty in the next test (line \lines{\ref{fig:uSPSC}}{63}), 
the algorithm fails because the first element pushed by the produces is definitely lost.
So, depending on the memory consistency model, we could have different scenarios.
%%\ma{Questo problema e' assimilabile a ABA? }
Consider a memory consistency model in which $W \rightarrow W$ program order
is respected. In this case, the emptiness check at line \lines{\ref{fig:uSPSC}}{63} could never
fail because a write in the internal SPSC buffer (line \lines{\ref{fig:SPSC}}{29}) cannot 
bypass the update of the buf\_w pointer (line \lines{\ref{fig:uSPSC}}{55}).
Instead, if $W \rightarrow W$ memory ordering is relaxed, the algorithm fails if the 
SPSC buffer has size 1, but it works if SPSC internal buffer has size greater than 1. 
In fact, if the SPSC internal buffer has size 1 it is possible that the write in the buffer is not seen
at line \lines{\ref{fig:uSPSC}}{63} because writes can be committed out-of-order in memory, and also, 
the Write Memory Barrier (WMB) at line \lines{\ref{fig:SPSC}}{28} is not sufficient, because it ensures 
that only the previous writes are committed in memory. 
On the other hand if the size of the SPSC buffer is at least 2 
the first of the 2 writes will be visible at line \lines{\ref{fig:uSPSC}}{63} because of the WMB 
instruction, thus the emptiness check could never fail.
From the above reasoning follows two theorems:

\begin{theorem}
The uSPSC queue is correct under any memory consistency
model that ensure $W \rightarrow W$ program order.
\end{theorem}

\begin{theorem}
The uSPSC queue is correct under any memory consistency
model provided that the size of the internal circular buffer is
greater than 1.
\end{theorem}

\section{Experiments}
\label{sec:exp}

\begin{figure}
\begin{center}
\includegraphics[width=1.0\linewidth]{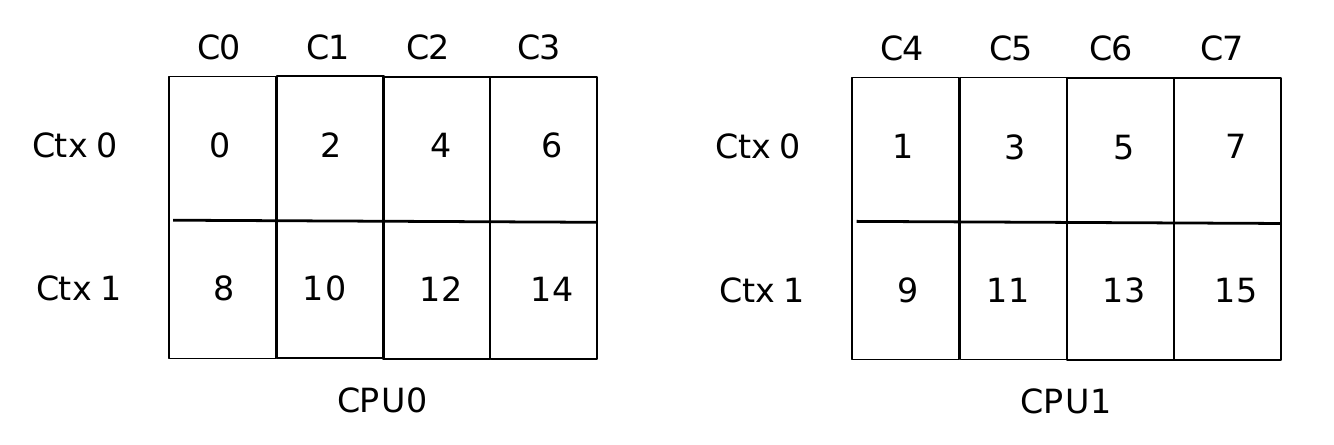}
\caption{
Core's topology on the Intel Xeon E5520 workstation used for the tests.\label{fig:topology}}
\end{center}
\end{figure}

All reported experiments have been executed on an Intel
workstation with 2 quad-core Xeon E5520 Nehalem (16 HyperThreads)
@2.26GHz with 8MB L3 cache and 24 GBytes of main memory with Linux x86\_64.
The Nehalem processor uses Simultaneous MultiThreading (SMT,
a.k.a. HyperThreading) with 2 contexts per core and the Quickpath
interconnect equipped with a distributed cache coherency protocol.
SMT technology makes a single physical processor appear as two logical
processors for the operating system, but all execution resources are
shared between the two contexts.
We have 2 CPUs each one with 4 physical cores. The operating system (Linux
kernel 2.6.20) sees the two per core contexts as two distinct cores 
assigning to each one a different id whose topology is sketched in 
Fig. \ref{fig:topology}.
%
%All presented experimental results are taken as an average of 5 runs exhibiting
%low variance.

The methodology used in this paper to evaluate performance consists in 
plotting the results obtained by running a simple synthetic benchmarks
and a very simple microkernel.

The first test is a 2-stage pipeline in which the first stage (P)
pushes 1 million tasks (a task is just a memory pointer) into a FIFO queue and 
the second stage (C) pops tasks from the queue and checks for correct values. Neither additional memory
operations nor additional computation in the producer or consumer stage 
is executed. With this simple test we are able to measure the raw performance
of a single push/pop operation by computing the average value of 100 runs and 
the standard deviation.
 
\begin{figure}
\begin{center}
\includegraphics[width=0.32\linewidth]{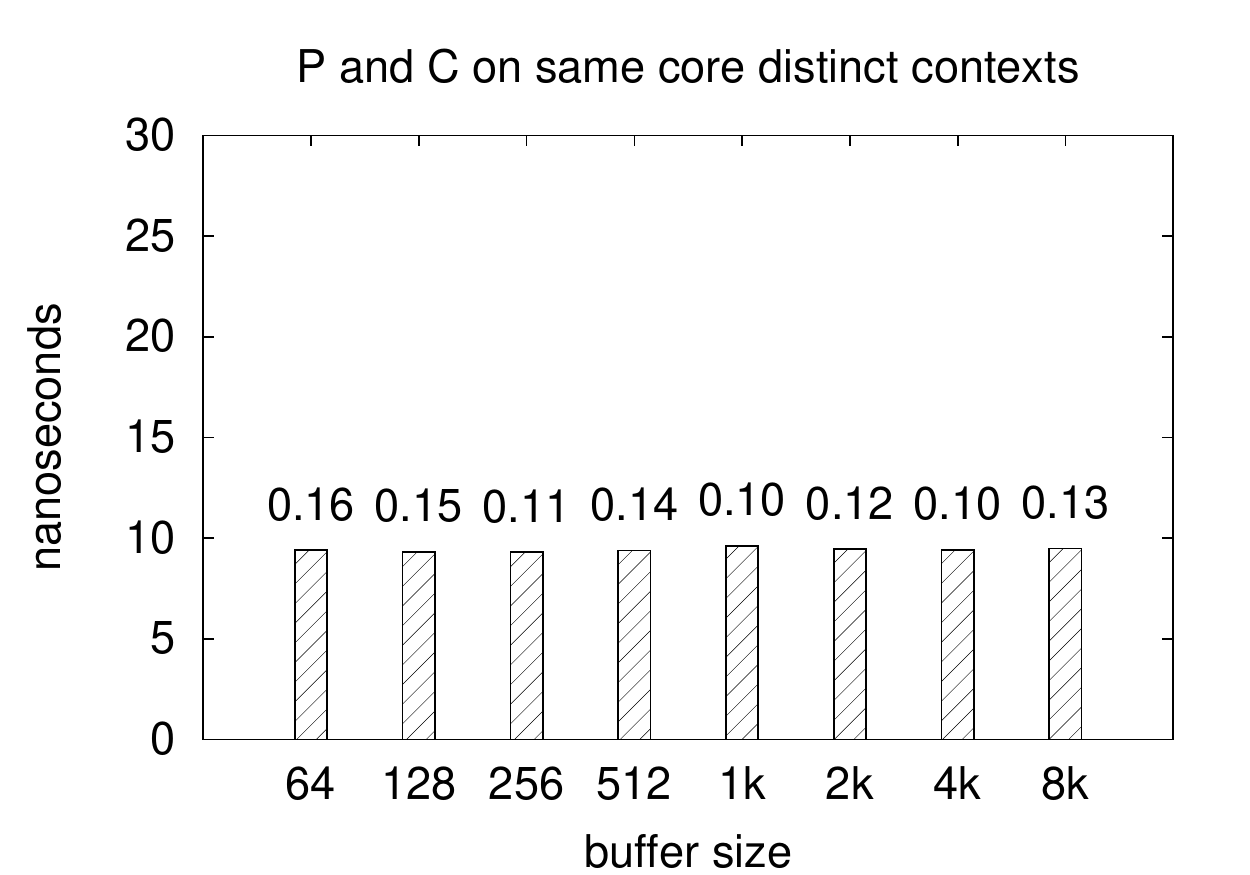}
\includegraphics[width=0.32\linewidth]{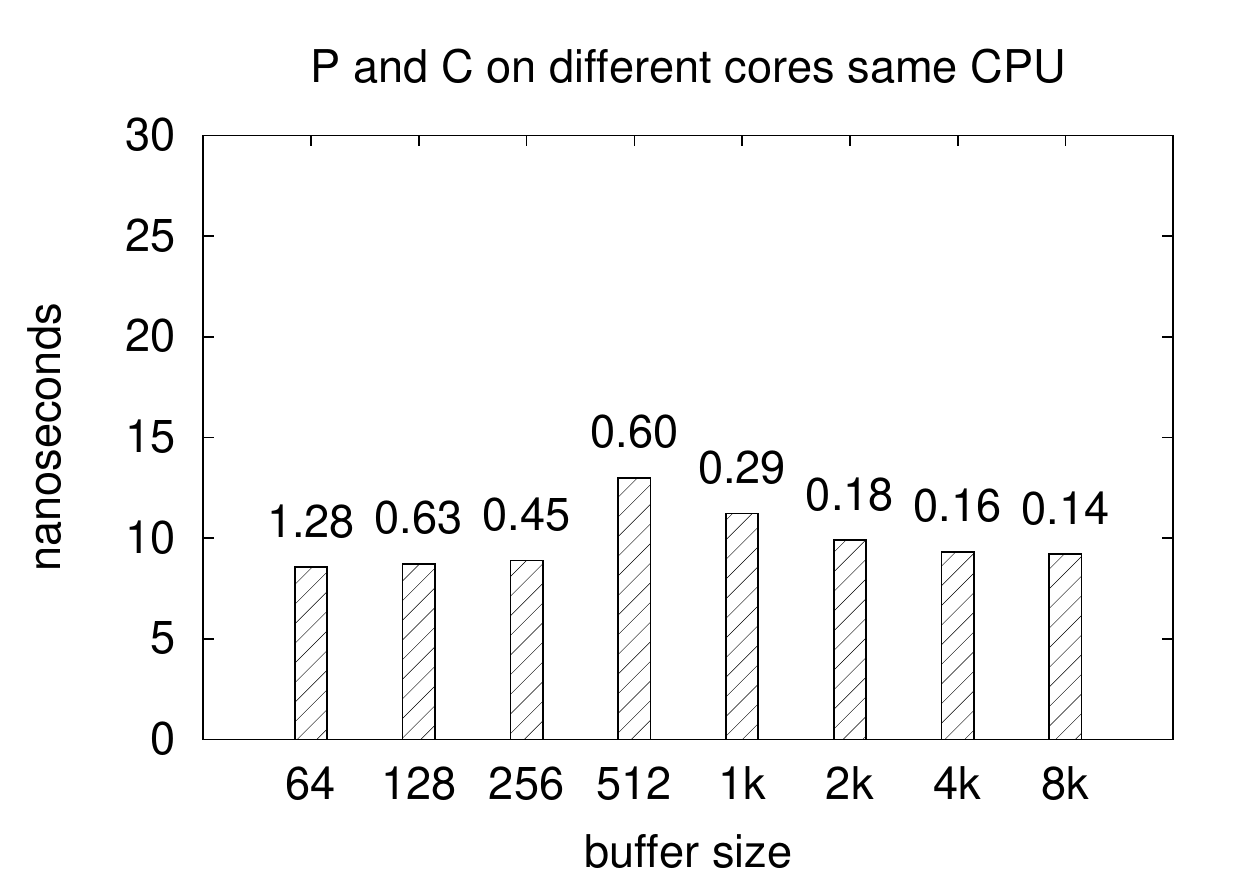}
\includegraphics[width=0.32\linewidth]{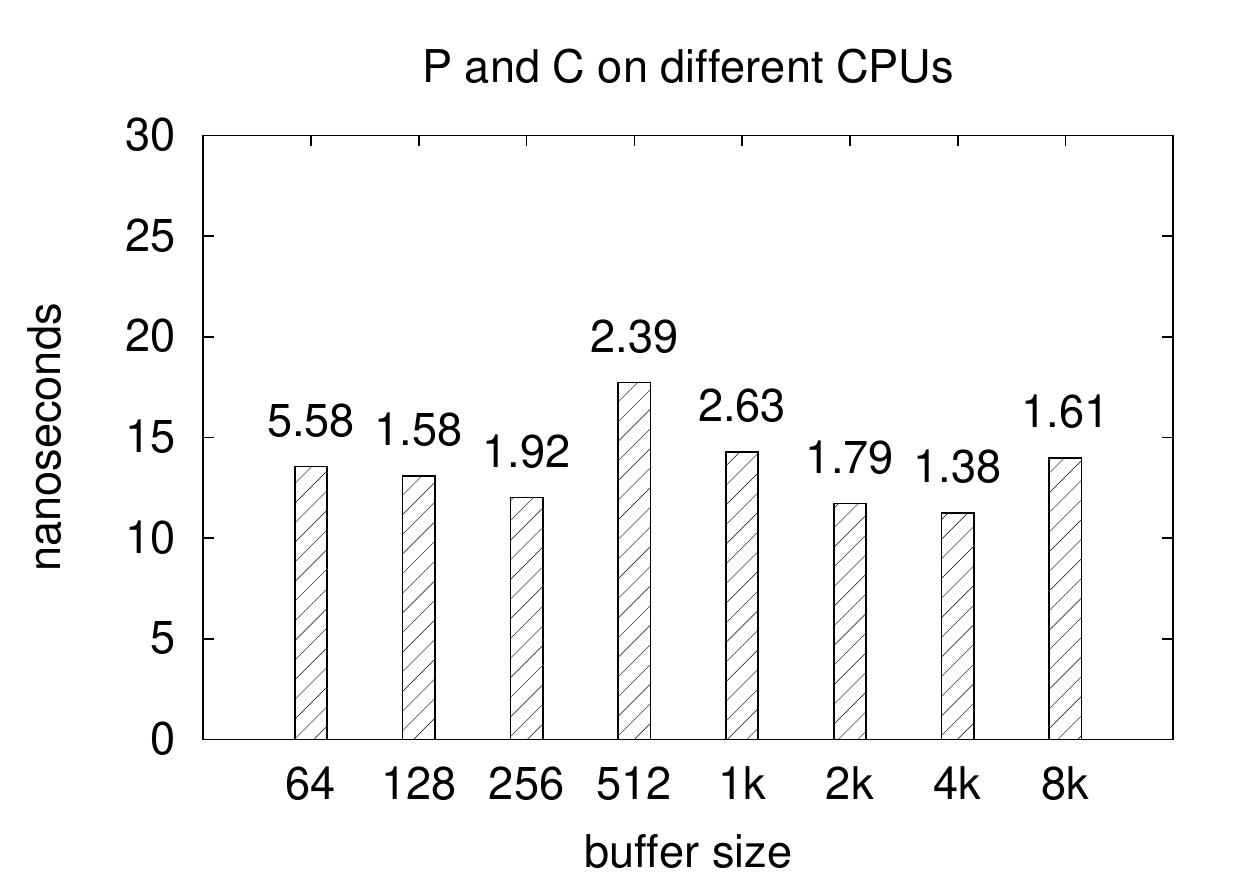}
\caption{
Average latency time and standard deviation (in nanoseconds) of a push/pop operation for the SPSC queue using different buffer size. The producer (P) and the consumer (C) are pinned: on the same core (left), on different core (middle), on different CPUs (right).\label{fig:SPSC:bounded}}
\end{center}
\end{figure}

\begin{figure}
\begin{center}
\includegraphics[width=0.32\linewidth]{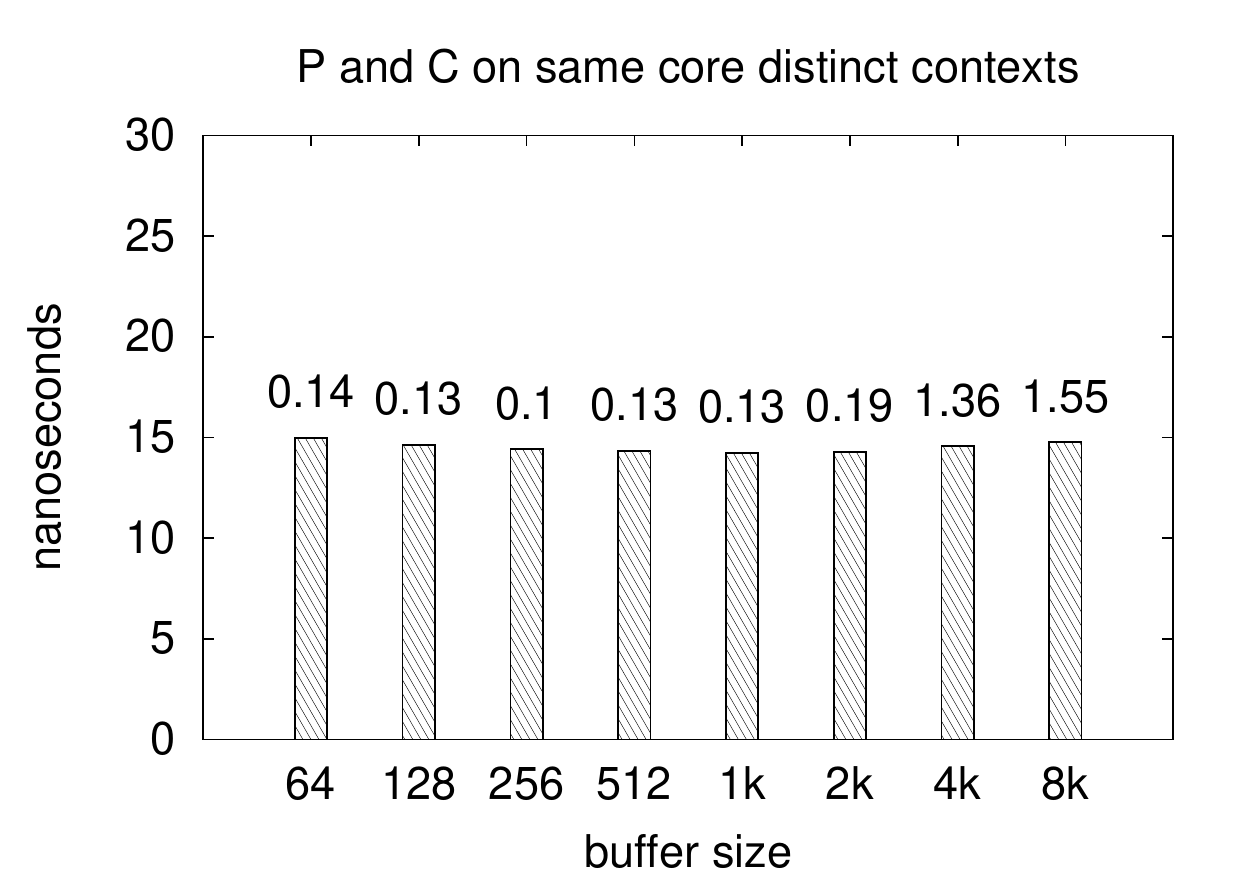}
\includegraphics[width=0.32\linewidth]{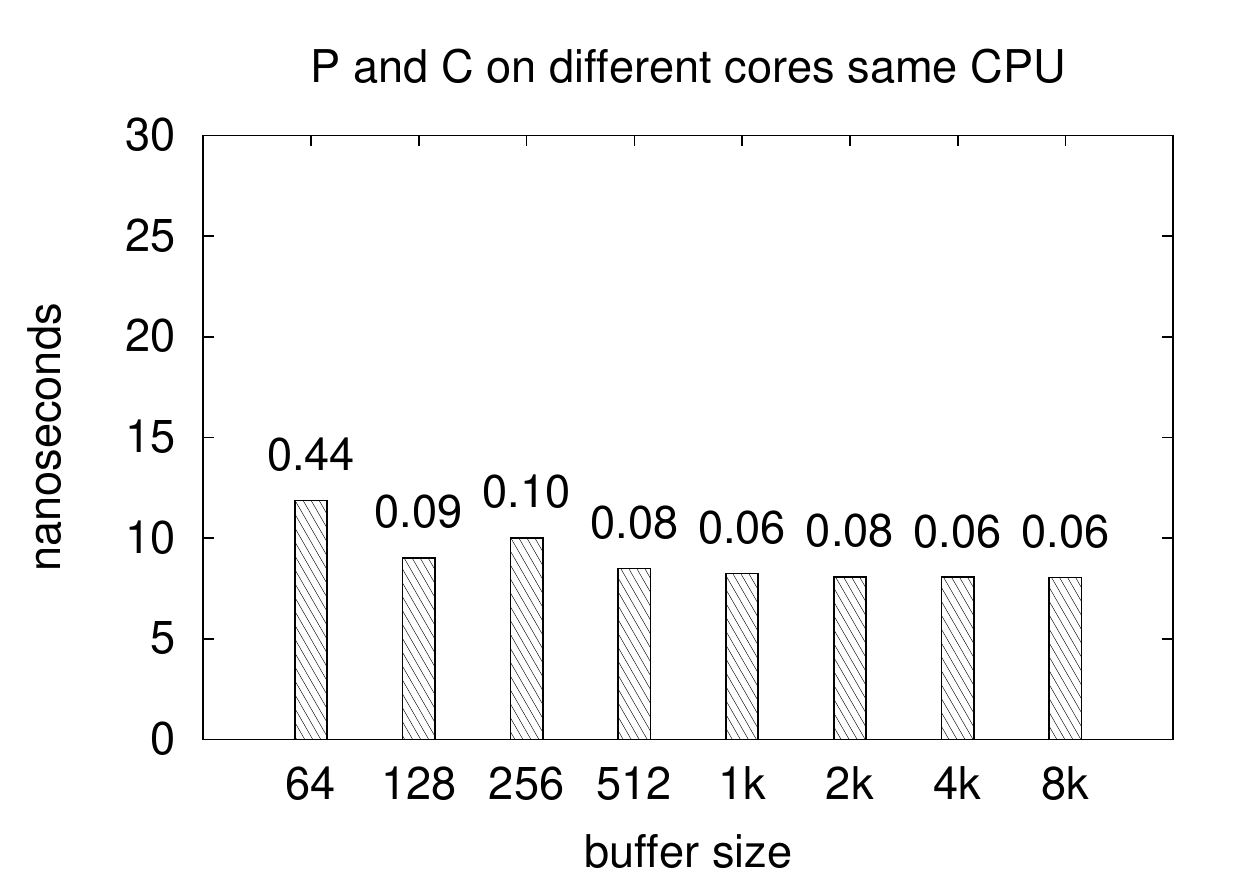}
\includegraphics[width=0.32\linewidth]{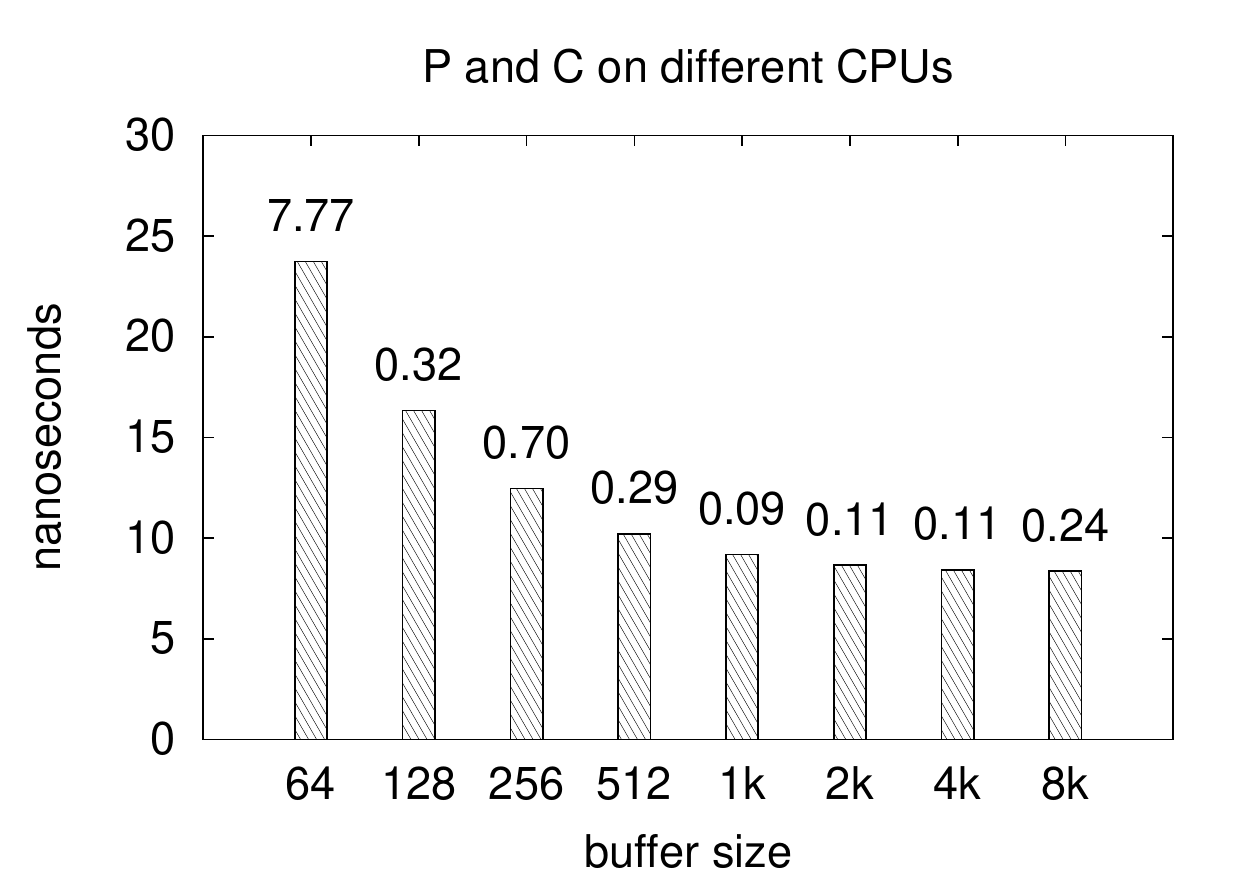}
\caption{
Average latency time and standard deviation (in nanoseconds) of a push/pop operation for the unbounded SPSC queue (uSPSC) using different internal buffer size. The producer (P) and the consumer (C) are pinned: on the same core (left), on different core (middle), on different CPUs (right).\label{fig:SPSC:unbounded}}
\end{center}
\end{figure}

\begin{figure}
\begin{center}
\includegraphics[width=0.32\linewidth]{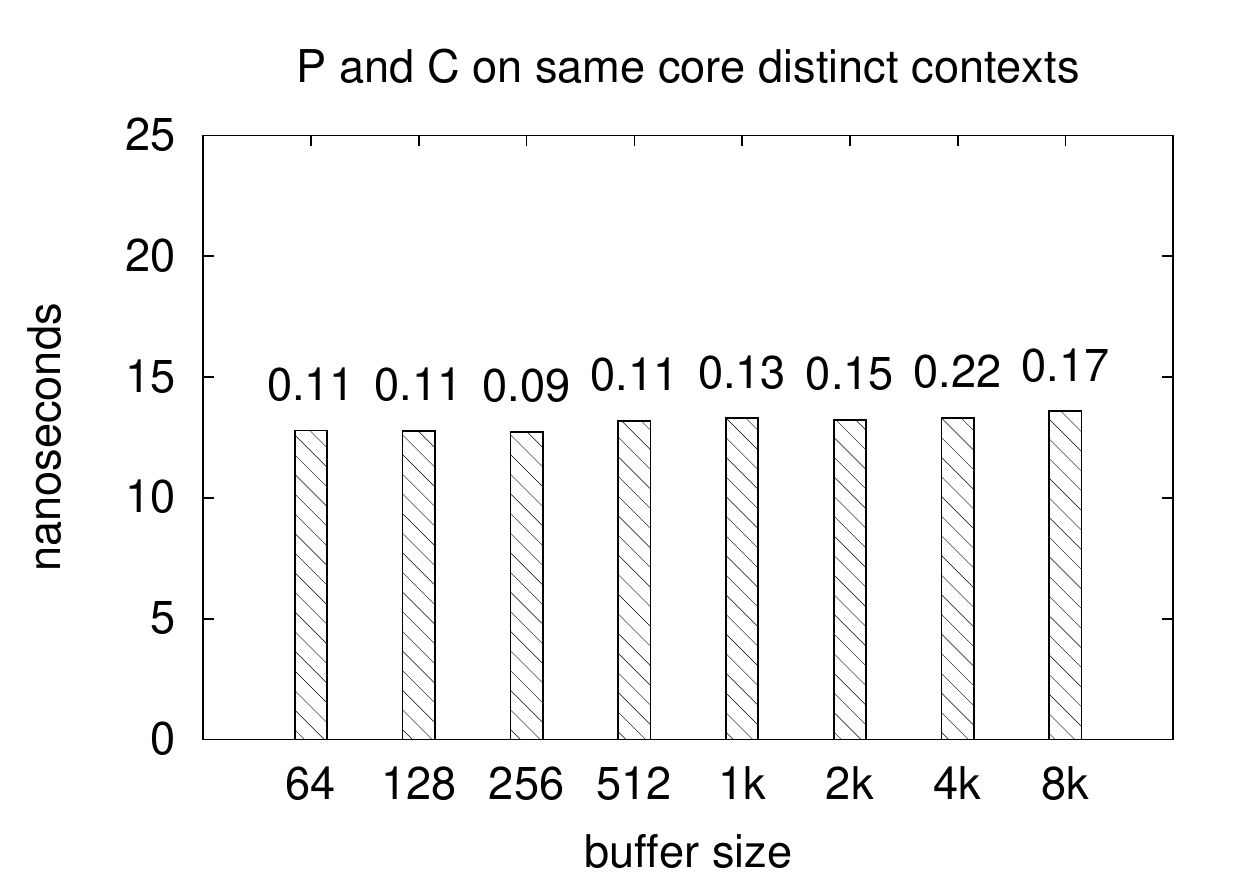}
\includegraphics[width=0.32\linewidth]{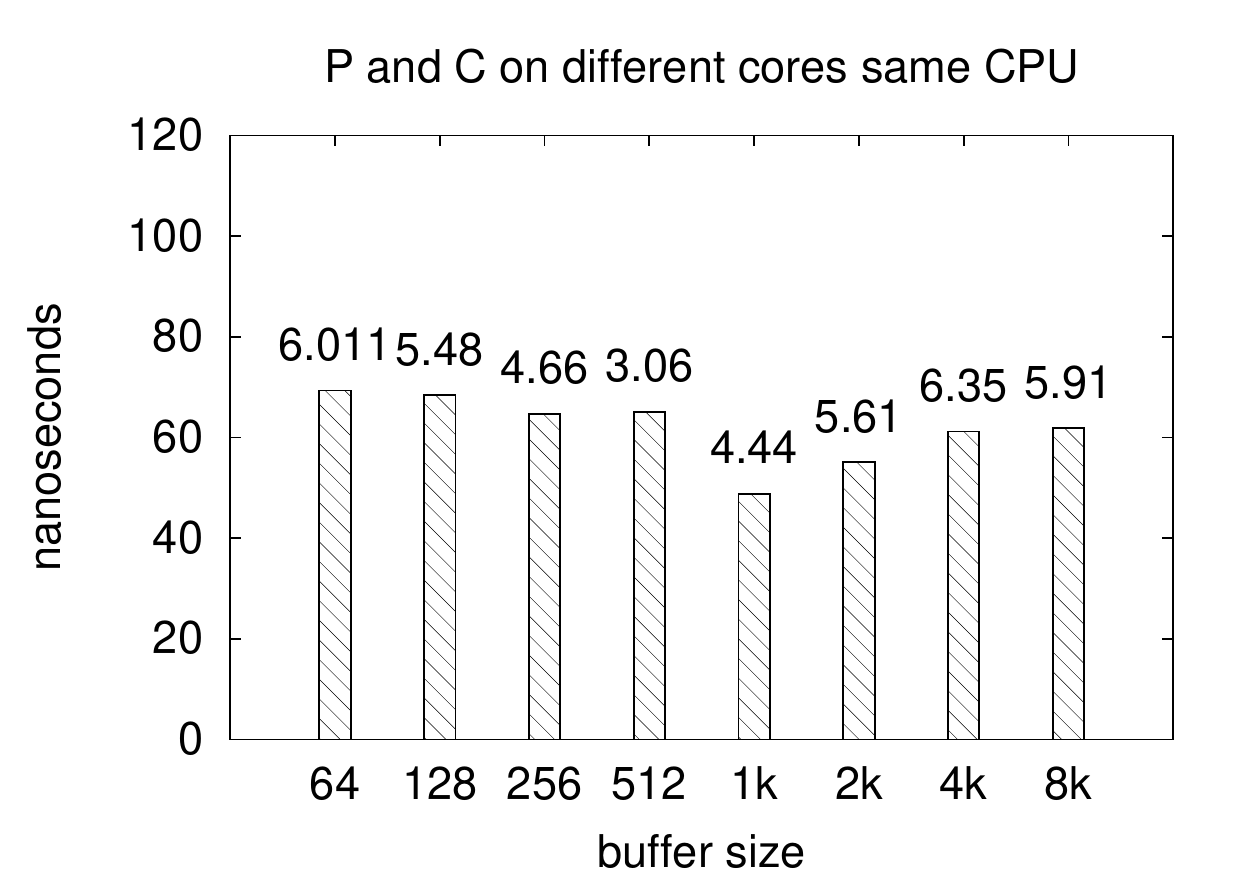}
\includegraphics[width=0.32\linewidth]{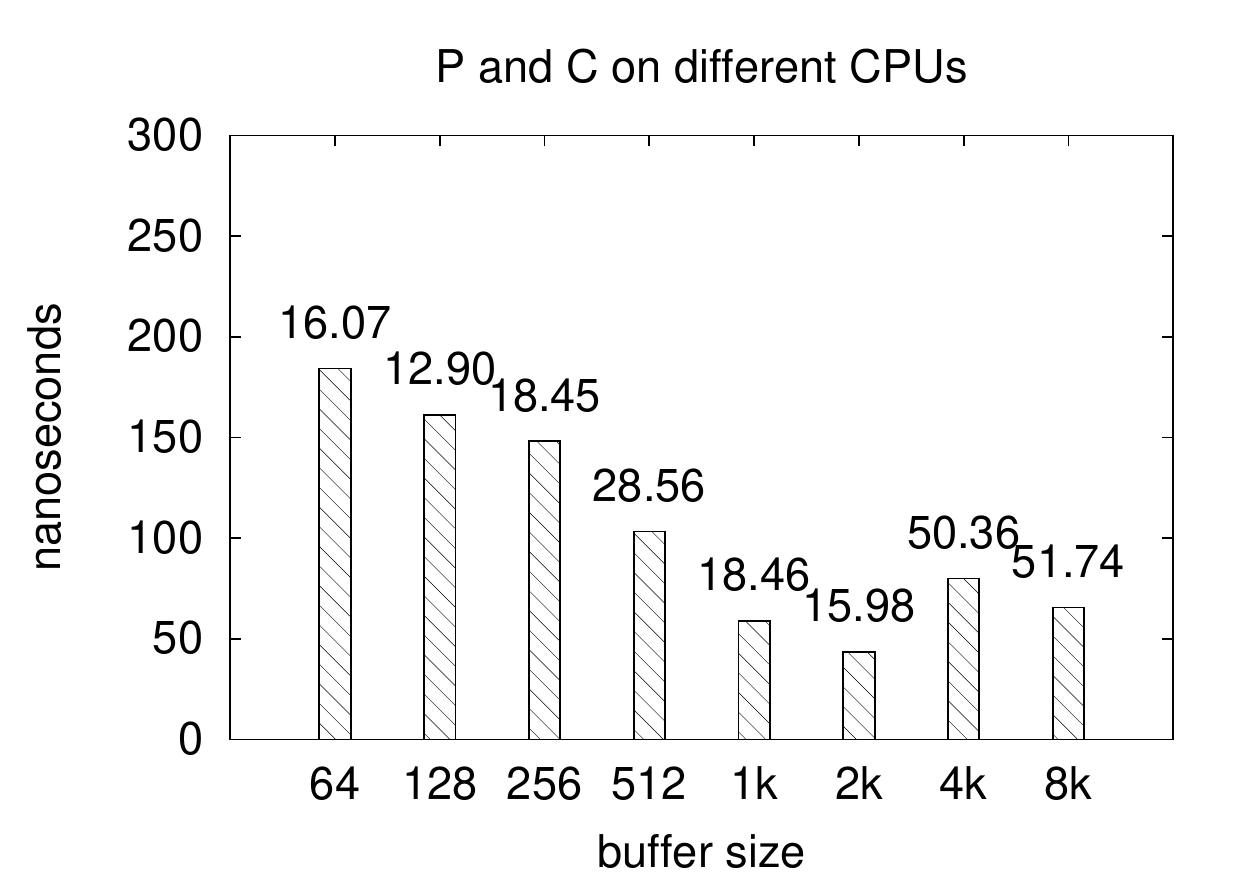}
\caption{
Average latency time and standard deviation (in nanoseconds) of a push/pop operation for the dynamic list-based SPSC queue (dSPSC) using different internal cache size. The producer (P) and the consumer (C) are pinned: on the same core (left), on different core (middle), on different CPUs (right).\label{fig:SPSC:dynamic}}
\end{center}
\end{figure}

In Fig. \ref{fig:SPSC:bounded} are reported the values obtained running the first benchmark for the 
SPSC queue, varying the buffer size. We tested 3 distinct cases obtained by changing the physical mapping 
of the 2 threads corresponding to the 2 stages of the pipeline: 
1) the first and second stage of the pipeline are pinned on the same physical core 
but on different HW contexts (P on core 0 and C on core 8), 2) are pinned on the same CPU but on different physical cores (P on core 0 and C on core 2), 
and 3) are pinned on two cores of two distinct CPUs (P on core 0 and C on core 1).
In Fig. \ref{fig:SPSC:unbounded} and in Fig. \ref{fig:SPSC:dynamic} are reported the 
values obtained running the same benchmark using the unbounded SPSC (uSPSC) queue and the 
dynamic list-based SPSC queue (dSPSC) respectively. On top of each bar is reported the standard deviation in nanoseconds computed over 100 runs.

The SPSC queue is insensitive to buffer size in all cases. It takes on average 10--12 ns per queue 
operation with standard deviations less than 1 ns when the producer and the consumer are on the same CPU, 
and takes on average 11--15 ns if the producer and the consumer are on separate CPUs. 
The unbounded SPSC queue (Fig. \ref{fig:SPSC:unbounded}) is more sensitive to the internal buffer size 
especially if the producer and the consumer are pinned into separate CPUs. The values obtained are 
extremely good if compared with the ones obtained for the dynamic list-based queue 
(Fig. \ref{fig:SPSC:dynamic}), and are almost the same if compared with the bounded SPSC queue
when using an internal buffer size greater than or equal to 512 entries.

The dynamic list-based SPSC queue is sensitive to the internal cache size (implemented with
a single SPSC queue). It is almost 6 times slower than the uSPSC version if the producer and the 
consumer are not pinned on the same core. In this case in fact, producer and consumer 
works in lock steps as they share the same ALUs and so dynamic memory allocation is reduced
with performance improvement. Another point in this respect, is that the dynamic list-based SPSC queue uses memory 
indirection to link together queue's elements thus not fully exploiting cache locality. The bigger the
internal cache the better performance is obtained. It is worth to note that caching strategies for dynamic
list-based SPSC queue implementation, significantly improve performance but are not enough 
to obtain optimal figures like those obtained in the SPSC implementation.

\begin{table}[tb]
\begin{tabular*}{\linewidth}{@{\hspace{4ex}}r@{\hspace{4ex}}rrrrr}
\toprule
\  & L1 accesses & L1 misses  & L2 accesses  & L2 misses \\
%%cmidrule(rl){2-5}
push    & 9,845,321  & 249,789 &  544,882  & 443,387 \\
mpush   & 4,927,934  & 148,011 &  367,129  & 265,509 \\
\bottomrule
\end{tabular*}
\caption{push vs. mpush cache miss obtained using a SPSC of size 1024 and performing 1 million push/pop operations. \label{tab:cachemiss}}
\end{table}

We want now to evaluate the benefit of the cache optimization presented in Sec. \ref{sec:cacheopt} for the SPSC and for the uSPSC queue. We refer to mSPSC and to muSPSC the version of the SPSC queue and of the uSPSC queue which use the mpush instead of the push method.
Table \ref{tab:cachemiss} reports the L1 and L2 cache accesses and misses for the \texttt{push} and \texttt{mpush} methods using a specific
buffer size. As can be noticed, the mpush method greatly reduce cache accesses and misses. 
The reduced number of misses, and accesses in general, leads to better overall performance. The average 
latency of a push/pop operation, decreases from 10--11ns of the SPSC queue, to 6--9ns for the multi-push version. 
The comparison of the \texttt{push} and \texttt{mpush} methods for both the SPSC and uSPSC queue, distinguishing the three mapping scenario for the producer and the consumer, are shown in Fig. \ref{fig:SPSC:multipush}. The muSPSC queue is less sensitive to 
the cache optimization introduced with the \texttt{mpush} method with respect to the uSPSC queue.

\begin{figure}
\begin{center}
\includegraphics[width=0.32\linewidth]{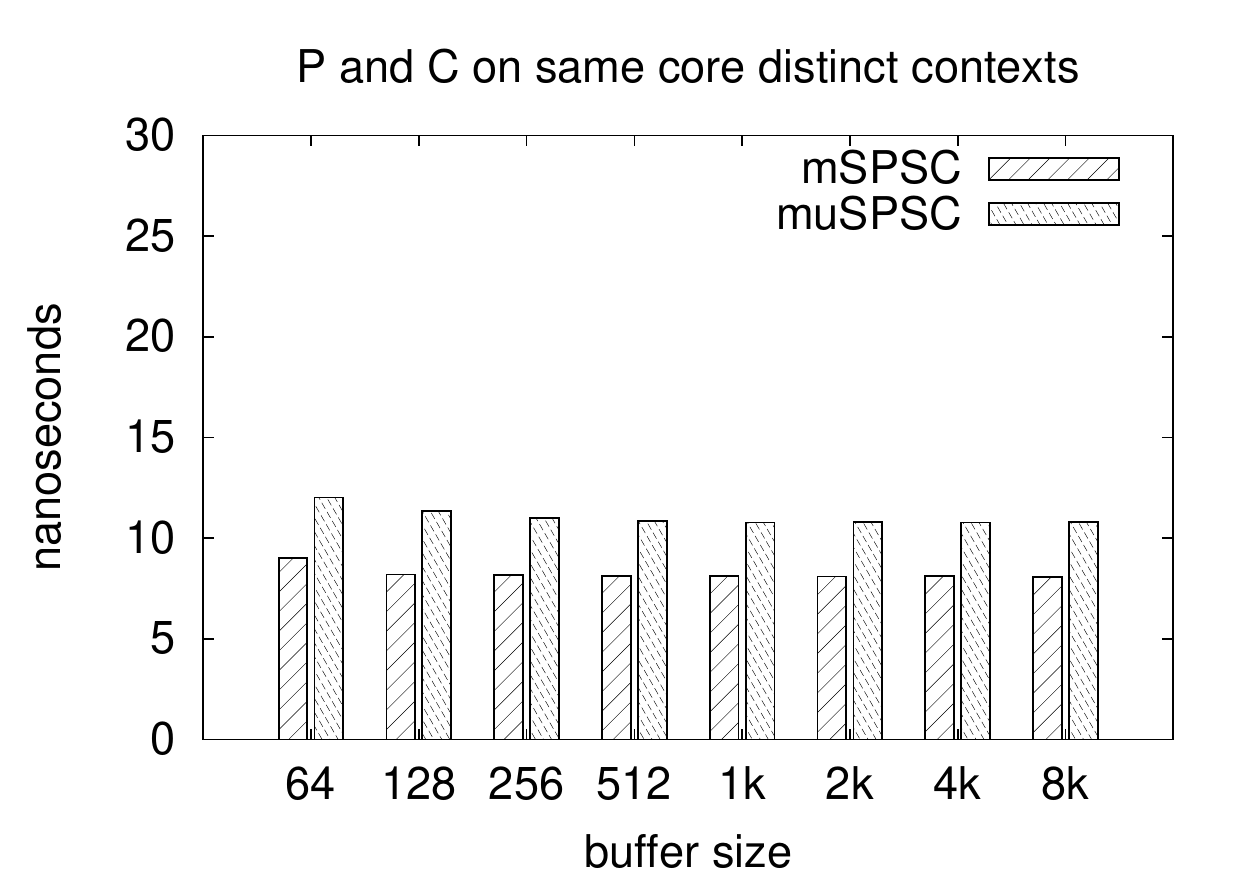}
\includegraphics[width=0.32\linewidth]{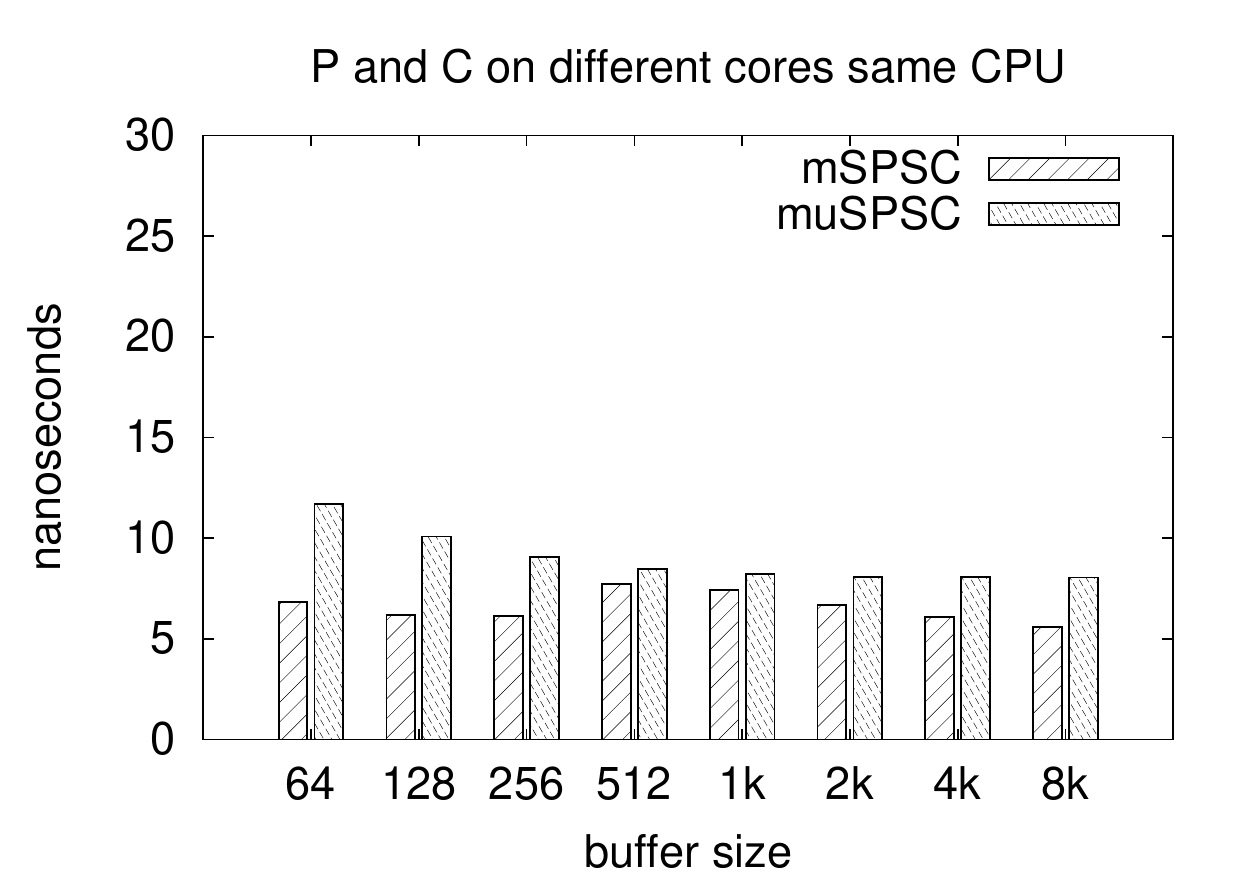}
\includegraphics[width=0.32\linewidth]{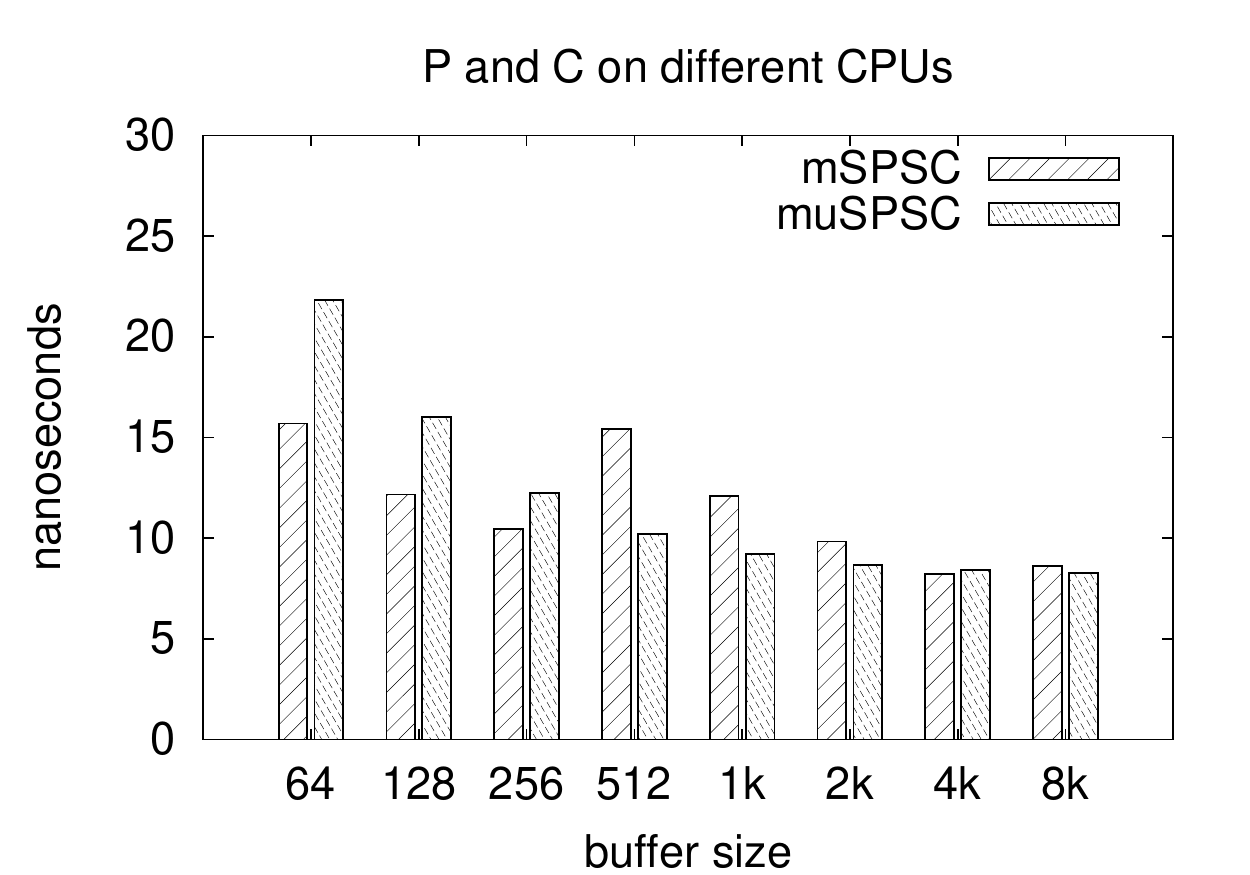}
\caption{
Average latency time of a multi-push/pop operation for the bounded and unbounded SPSC buffer. The multi-push internal buffer size is statically set to 16 entries. The producer (P) and the consumer (C) are pinned: on the same core (left), on different core (middle), on different CPUs (right).\label{fig:SPSC:multipush}}
\end{center}
\end{figure}

\begin{figure}
\begin{Bench}{}{}
int main() {
  double x = 0.12345678, y=0.654321012;
  for(int i=0;i<1000000;++i) {
    x = 3.1415 * sin(x);
    y += x - cos(y);
  }
  return 0;
}
\end{Bench}
\caption{Microbenchmark: sequential code.\label{fig:microbench}}
%%\hspace{0.5ex}
\end{figure}

\begin{figure}ls
\begin{Bench2}{}{}
void P() {
  double x = 0.12345678;
  for(int i=0;i<1000000;++i) {
    x = 3.1415 * sin(x);
    Q.push(x);
  }
}
void C() {
  double x, y=0.654321012;
  for(int i=0;i<1000000;++i) {
    Q.pop(&x);
    y += x - cos(y);
  }
}
\end{Bench2}
\caption{Microbenchmark: pipeline implementation.\label{fig:microbenchPipe}}
%%\hspace{0.5ex}
\end{figure}

\begin{figure}
\begin{center}
\includegraphics[width=0.32\linewidth]{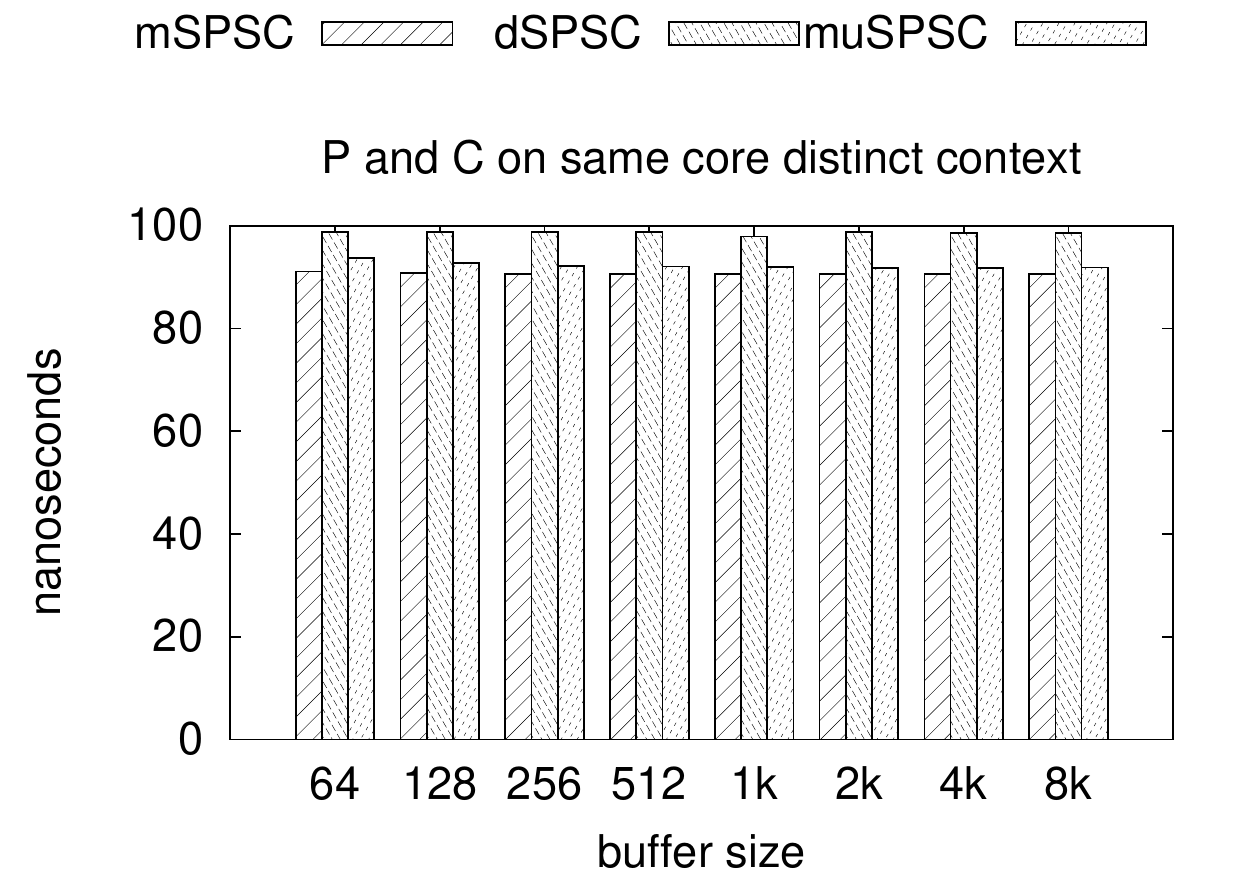}
\includegraphics[width=0.32\linewidth]{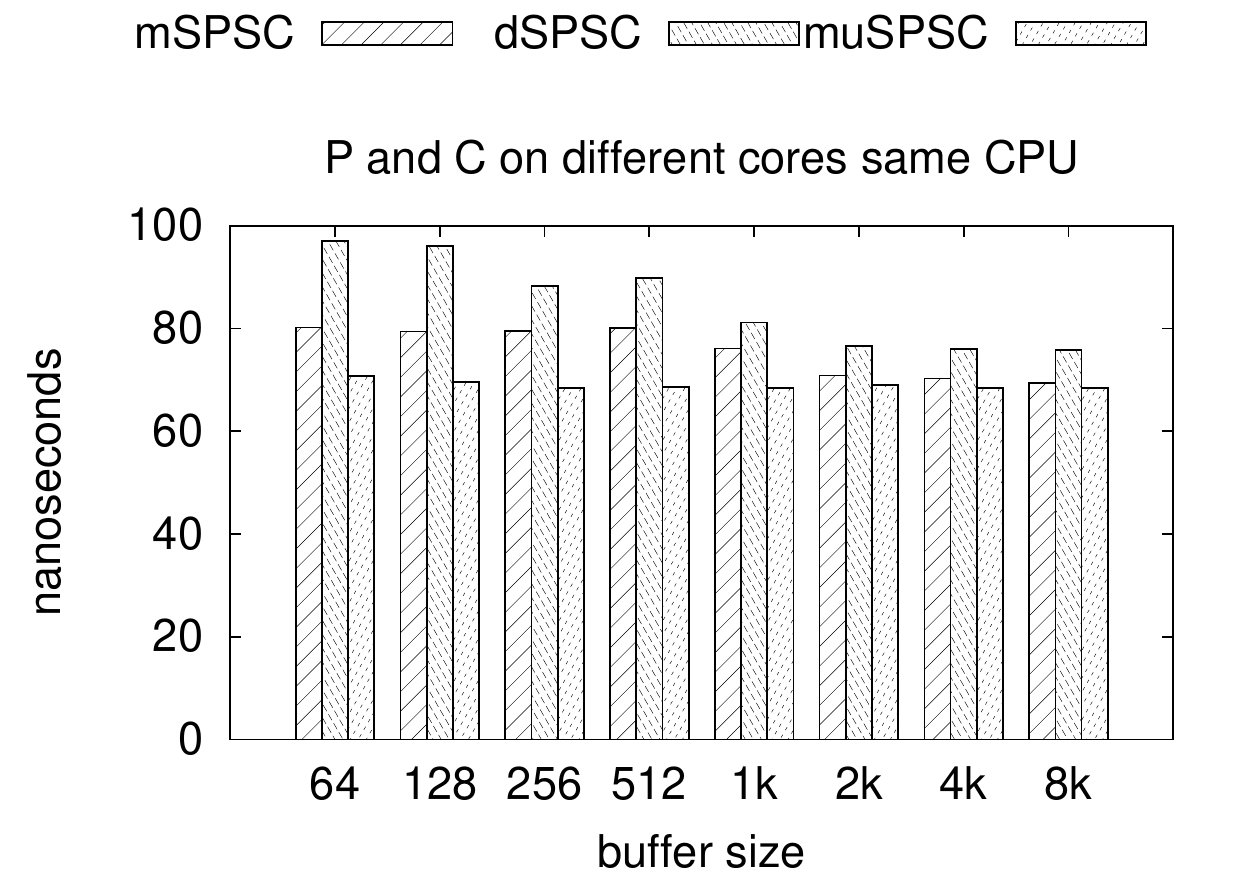}
\includegraphics[width=0.32\linewidth]{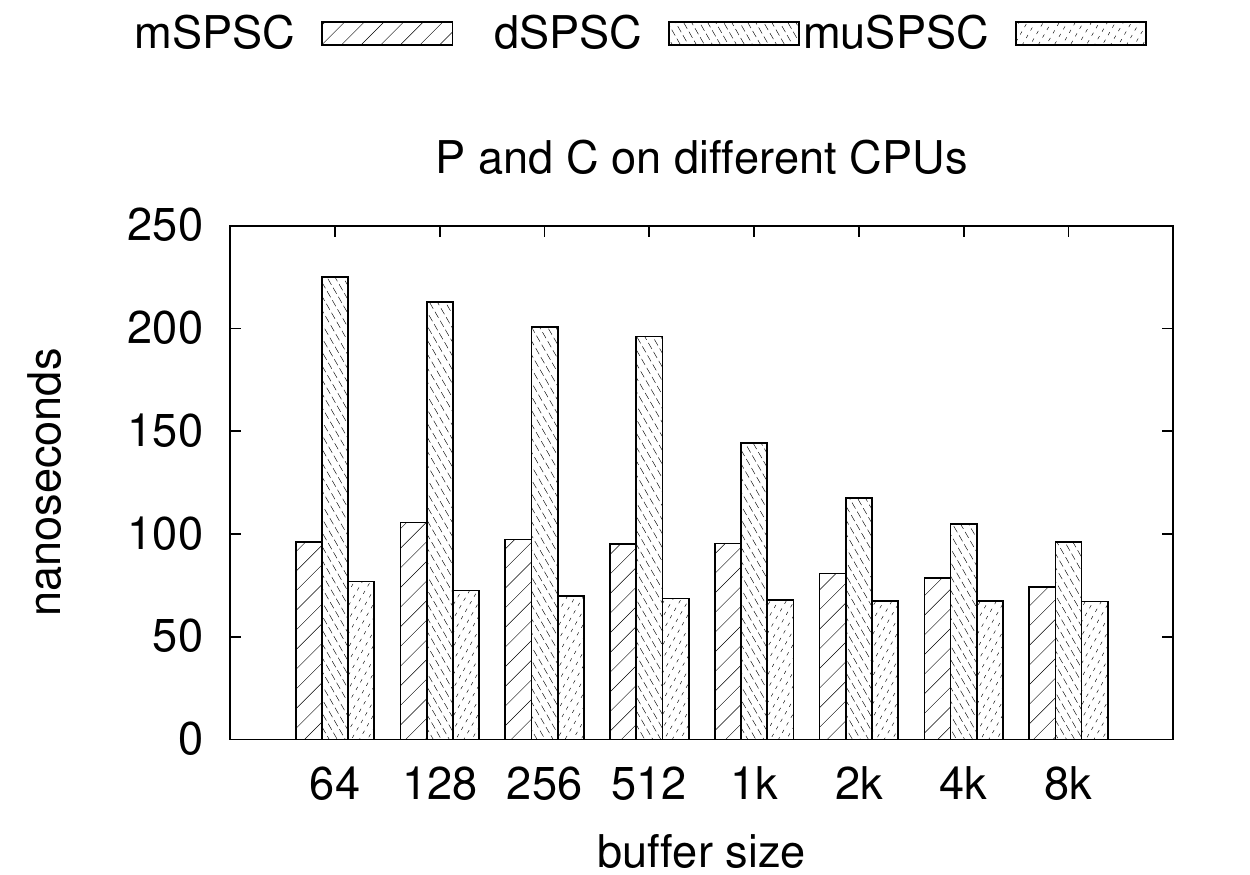}
\caption{Average latency time (in nanoseconds) of the pipeline microbenchmark implementation when using the mSPSC, dSPSC and muSPSC queue. P and C are pinned: on the same core (left), on different core (middle), on different CPUs (right).\label{fig:microbenchSPSC}}
\end{center}
\end{figure}

In order to test a simple but real application kernel we consider the code in Fig. \ref{fig:microbench}. 
The sequential execution of such code on a single core of the tested architecture is 94.6ms. We 
parallelize the code into a pipeline of 2 stages, P and C, as shown in Fig. \ref{fig:microbenchPipe}.
The 2 stages are connected by a FIFO queue. The results obtained considering for the queue the mSPSC, 
dSPSC and muSPSC implementations are shown in Fig \ref{fig:microbenchSPSC}. As can be noticed
the unbounded multi-push implementation (muSPSC) obtain the best performance reaching a maximum speedup of 
1.4, whereas the bounded multi-push implementation (mSPSC) reaches a maximum speedup of 1.28 and finally the
dynamic list-based queue (dSPSC) does not obtain any performance improvement reaching a maximum speedup 
of 0.98. This simple test, proves the effectiveness of the uSPSC queue implementation with respect to 
the list-based FIFO queue implementation when used in real case scenario.

\section{Conclusions} 
\label{sec:conclusions}
In this report we reviewed several possible implementations of fast wait-free 
Single-Producer/Single-Consumer (SPSC) queue for shared cache multi-core starting from
the well-known Lamport's circular buffer algorithm.
A novel implementation of unbounded wait-free SPSC queue has been introduced with a formal proof of
correctness. The new implementation is able to minimize dynamic memory allocation/deallocation
and increases cache locality thus obtaining very good performance figures on modern shared cache 
multi-core. We believe that the unbounded SPSC algorithm presented here can be used as an efficient alternative to 
the widely used list-based FIFO queue.

\section*{Acknoweledments}
The author gratefully acknowledge Marco Aldinucci, Marco Danelutto,
Massimiliano Meneghin and Peter Kilpatrick for their comments and suggestions.

\bibliographystyle{abbrv}
%%\bibliography{UniPisaGroup,ac,multicore}

\end{document}